\documentclass[prd
,preprint
,showkeys
,showpacs
,nofootinbib,
,amsfonts,amsmath,amssymb
]{revtex4}
\usepackage[pdftex]{graphicx}
\usepackage{float}
\usepackage{color}
\newcommand{\bs}[1]{\boldsymbol{#1}}
\newcommand{\pa}{\partial}
\newcommand{\al}{\alpha}
\newcommand{\del}{\delta}

\begin{document}
\title{Harvesting large scale entanglement in de Sitter space \\
  with multiple detectors}
\author{Shingo Kukita}
\email{kukita@th.phys.nagoya-u.ac.jp}
\author{Yasusada Nambu}
\email{nambu@gravity.phys.nagoya-u.ac.jp}
\affiliation{Department of Physics, Graduate School of Science, Nagoya 
University, Chikusa, Nagoya 464-8602, Japan}

\date{August 29, 2017 ver 1.0} 
\begin{abstract}
  We consider entanglement harvesting in de Sitter space using a model
  of multiple qubit detectors. We obtain the formula of the
  entanglement negativity for this system. Applying the obtained
  formula, we find that it is possible to access to the entanglement
  on the super horizon scale if sufficiently large number of detectors
  are prepared. This result indicates the effect of the multipartite
  entanglement is crucial for detection of large scale entanglement in
  de Sitter space.
\end{abstract}
\maketitle

\section{Introduction}

All structure in the Universe can be traced back to primordial quantum
fluctuations generated during an inflationary phase of the very early
universe. To apprehend history and origin of our universe, it is
essential to understand the mechanism and the nature of these
fluctuation with quantum origin. To investigate the quantum property
of primordial quantum fluctuation, the entanglement is a key concept
to distinguish the quantum nature from the classical one. Thus, it is
an important task to analyze detail of the entanglement of quantum
fluctuations generated by inflation.  In this direction, detection of
the entanglement of the quantum scalar field using a pair of particle
detectors were
considered~\cite{Steeg2009,nambu2011,Martin-Martinez2012,Nambu2013,Kukita2017}. The
entanglement  of the scalar field can be probed by evaluating the
entanglement between these two detectors interacting with the field;
An initially non-entangled pair of detectors can evolve to be an
entangled state through the interaction with the quantum field. As the
entanglement cannot be created by local operations, this implies that
the entanglement of the quantum field is transferred to the pair of
detectors.

In de Sitter spacetime, although detectors can probe the entanglement
of the scalar field on the scale smaller than the Hubble horizon, they
cannot catch entanglement beyond the Hubble horizon
scale~\cite{Steeg2009,nambu2011,Nambu2013,Kukita2017}. Qualitatively
similar result is shown for entanglement between two spatial regions
defined via averaging (coarse graining of the scalar
field)~\cite{Nambu2008,Nambu2009}, and connection to the quantum to
classical transition in the early universe was discussed.  On the
other hand, the result of recent lattice
calculation~\cite{Matsumura2017a} which simulates the quantum scalar
field near the continuous limit shows entanglement between two spatial
regions persists even beyond the Hubble horizon scale and the
entanglement negativity does not vanish.  This discrepancy may come
from efficiency of entanglement detection using a pair of detectors;
we expect that the efficiency of detection increases if the degrees of
freedom of detectors grows. Hence, we consider multiple detectors
system and investigate how the maximum possible distance of
entanglement detection depends on the number of detectors.

In this paper, we consider $m+n$ qubit detectors and investigate
detectability of bipartite entanglement on the super horizon scale in
de Sitter space.  We obtain negativity of this system analytically in
the lowest non-trivial order of perturbation with respect to the
coupling constant. Using this result, we discuss possibility of
entanglement harvesting beyond the Hubble horizon scale.  The
structure of the paper is as follows. In Sec.~II, we introduce our
model of multiple detectors system and the master equation for
detectors state. In Sec.~III, we review two quits detectors case and show the
maximum possible distance of entanglement detection cannot exceeds the
Hubble horizon scale. In Sec.~IV, we evaluate the negativity of $m+n$
detectors case. In Sec.~V, we discuss the relation to the monogamy
inequality. Sec.~VI is devoted to summary. We use the unit in which
$\hbar=c=1$ throughout the paper.

\section{Model and strategy}

\begin{figure}[H]
  \centering
  \includegraphics[width=0.38\linewidth,clip]{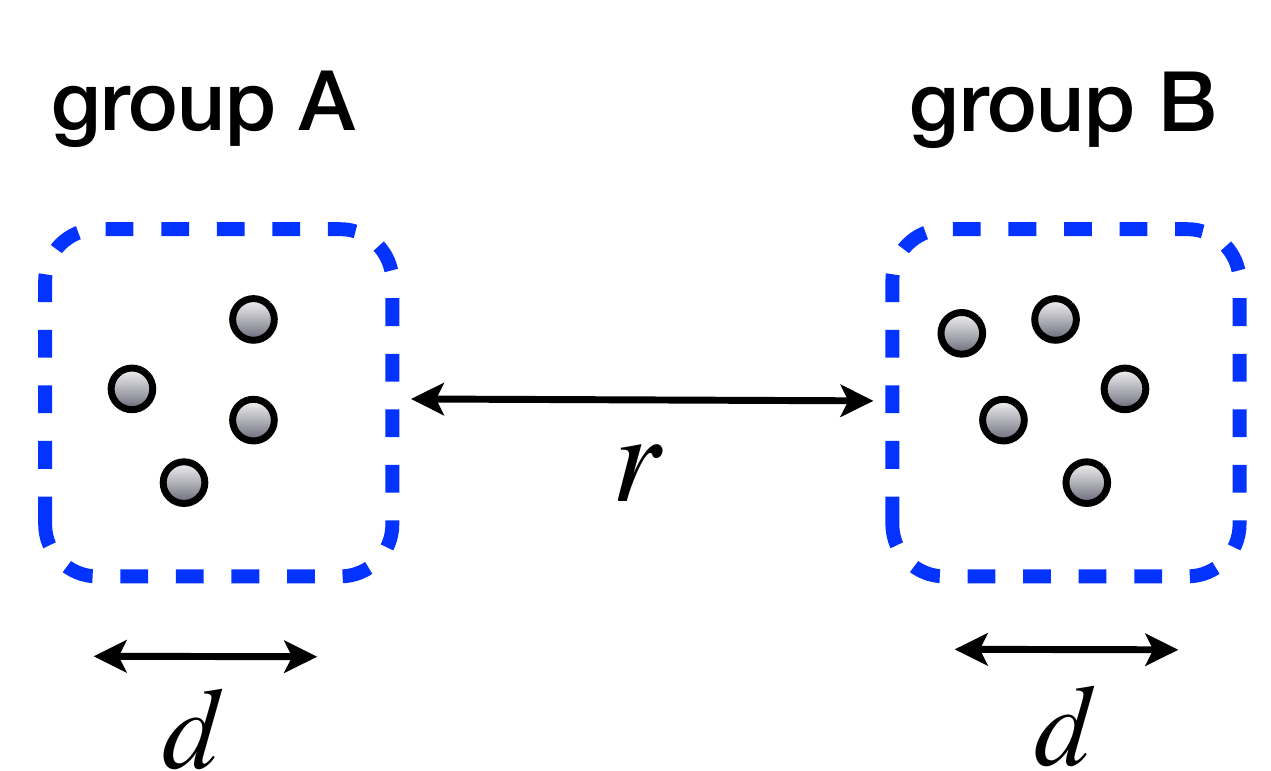}
  \caption{ Setup of $m+n$ detectors system. The group A consists of
    $m$ detectors and the group B consists of $n$ detectors. The
    separation between two groups is $r$ and size of each group is
    $d$. We assume $d\leq r$.}
  \label{fig:setup}
\end{figure}

We consider the following Hamiltonian for $m+n$ qubit detectors
interacting with a scalar field $\phi$ (see Fig.~\ref{fig:setup}):
\begin{equation}
  H_{\text{tot}}=H_S+H_\text{int}+H_\phi=
\sum_{\al=1}^{m+n}\frac{\omega}{2}\,\sigma_3^{(\al)}+g
\sum_{\al=1}^{m+n}(\sigma_{+}^{(\al)}+\sigma_{-}^{(\al)})\,\phi(\bs{x}_\al)+H_\phi,
\end{equation}
where $\omega$ represents energy difference between two internal levels
$|0\rangle, |1\rangle$ and $g$ is a coupling constant between detectors
and the scalar field. The tensor products of the operators are
defined as
\begin{equation}
  \sigma_j^{(1)}=\sigma_j\otimes\openone\otimes\cdots\otimes\openone,\quad
\sigma_j^{(2)}=\openone\otimes\sigma_j\otimes\openone\otimes\cdots\otimes\openone,\quad
\cdots,
\sigma_j^{(m+n)}=\openone\otimes\openone\otimes\cdots\otimes\openone
\otimes\sigma_j,
\end{equation}
with $\sigma_3=|1\rangle\langle 1|-|0\rangle\langle 0|,
\sigma_{+}=|1\rangle\langle 0|, \sigma_{-}=|0\rangle\langle 1|$.
As a tool of our analysis, we introduce a master equation for
detectors' state.  Regarding the scalar field as an environment, we
obtain the reduced density matrix for detectors by tracing out the
scalar field degrees of freedom. Provided that the time scale of the
environment is shorter than the detector's time scale and the coupling
is weak $g\ll 1$, the state of detectors $\rho$ can be shown to obey
the following Gorni-Kossakowski-Lindblad-Sudarshan (GKLS) type master
equation~\cite{Lidar2001,Schaller2008,Benatti2010,Majenz2013,Nambu2016a}:
\begin{align}
  &\frac{\pa\rho}{\pa t}+i[H_{\text{eff}},\rho]=\mathcal{L}[\rho],
  \label{eq:master}\\ 
  &H_\text{eff}=H_S-\frac{i}{2}\sum_{\al_1,\al_2=1}^{m+n}\sum_{j_1,j_2=\pm}H_{j_1j_2}
    ^{(\al_1\al_2)}\sigma_{j_1}^{(\al_1)}\sigma_{j_2}^{(\al_2)},
\\
  &\mathcal{L}[\rho]=\frac{1}{2}\sum_{\al_1,\al_2=1}^{m+n}\sum_{j_1,j_2=\pm}
    C_{j_1j_2}^{(\al_1\al_2)}\left[2\sigma_{j_2}^{(\al_2)}\rho\,\sigma_{j_1}^{(\al_1)}
    -\sigma_{j_1}^{(\al_1)}\sigma_{j_2}^{(\al_2)}\rho-\rho\,\sigma_{j_1}^{(\al_1)}
    \sigma_{j_2}^{(\al_2)}\right],
\end{align}
where $H_\text{eff}$ is the effective Hamiltonian of the detectors
system with quantum corrections. The coefficients
$H_{j_1\!j_2}^{(\al_1\al_2)}, C_{j_1\!j_2}^{(\al_1\al_2)}$ are
expressed using the Wightman function of the quantum field as
\begin{align}
  &C_{j_1j_2}^{(\al_1\al_2)}=\frac{2g^2}{\pi\sigma}e^{-\omega^2\sigma^2}
e^{i\omega\sigma j_{+}}
    \!\!\int_{-\infty}^{+\infty}\!\!dx\,dy\,e^{-\frac{1}{\sigma^2}\left[x-\left(\sigma+\frac{i}{2}\omega\sigma^2j_{+}\right)\right]^2-\frac{1}{\sigma^2}\left(y-\frac{i}{2}\omega\sigma^2j_{-}\right)^2}  
D(r_c,t+x,y),
  \\
  &H_{j_1j_2}^{(\al_1\al_2)}=\frac{2g^2}{\pi\sigma}e^{-\omega^2\sigma^2}
e^{i\omega\sigma j_{+}}
   \!\! \int_{-\infty}^{+\infty}\!\!dx\,dy\,\mathrm{sgn}(y)
\,e^{-\frac{1}{\sigma^2}\left[x-\left(\sigma+\frac{i}{2}\omega\sigma^2j_{+}\right)\right]^2-\frac{1}{\sigma^2}\left(y-\frac{i}{2}\omega\sigma^2j_{-}\right)^2}
D(r_c,t+x,y),
\end{align}
with $j_{\pm}=j_1\pm j_2$ and $D(r_c,x,y)$ is the Wightman function of
the scalar field
\begin{equation}
  D(r_c,x,y)=\langle\phi(t_1,\bs{x}_{\al_1})\phi(t_2,\bs{x}_{\al_2})\rangle,\quad
  x=(t_1+t_2)/2,\quad y=(t_1-t_2)/2, 
\end{equation}
where $r_c=|\bs{x}_{\al_1}-\bs{x}_{\al_2}|$ denotes comoving distance
between two detectors. The parameter $\sigma$ in
$C_{j_1j_2}^{(\al_1\al_2)}, H_{j_1j_2}^{(\al_1\al_2)}$ specifies the
time scale of coarse graining which is necessary to derive the GKLS
master equation \eqref{eq:master}. In the limit of
$\sigma\rightarrow\infty$, this master equation reduces to that with
the rotating wave approximation which neglects transition via energy
nonconserving processes. The GKLS master equation preserves the trace
and complete positivity.

The master equation \eqref{eq:master} was applied to two detectors
system in de Sitter space for the purpose of investigating long time
evolution of negativity beyond the Hubble time
scale~\cite{Kukita2017}. In the analysis of the present paper, we
concentrate on short time evolution from the initial state and do not
solve this equation exactly. In such a restricted situation, as we
will show in Sec.~III, prediction by the master equation
\eqref{eq:master} coincides with that of the detectors with finite
interaction time which is usually imposed by introducing an
appropriate switching function of detector.

To examine detection of entanglement of the quantum field, we consider
a solution of the master equation \eqref{eq:master} with a separable
initial condition and judge the separability of the detectors state
after evolution. For $\Delta t=t-t_0 \ll 1/\omega$, the solution with
the initial state $\rho_0=\rho(t_0)$ is
\begin{equation}
  \rho(t)=\rho_0+\Delta t\left(-i[H_\text{eff},\rho_0]+\mathcal{L}[\rho_0]
\right). 
\label{eq:rhostate}
\end{equation}
For the initial separable state of detectors
$\rho_0=|0\cdots 0\rangle\langle 0\cdots 0|$, $H_\text{eff}$ and
$\rho_0$ commutes each other and the state after evolution can be
written as
\begin{equation}
    \rho(t)=\rho_0+\Delta t\mathcal{L}[\rho_0].
    \label{eq:sol}
\end{equation}
Thus the entanglement of the state $\rho$ is completely determined
only by the operator $\mathcal{L}[\rho_0]$. From now on, we examine
the state \eqref{eq:sol}.

We divide $m+n$ detectors to two groups and assign labels of detectors
as
\begin{equation}
  \text{A:}\quad \al\in 1,\cdots,m,\qquad\text{B:}\quad \al\in
  m+1,\cdots m+n.
\end{equation}
For simplicity of analysis, we assume distance between two groups is $r$, and
distance between two detectors belonging to the same group is $d$
(see Fig.~\ref{fig:setup}).
We denote possible states of $m+n$ detectors after evolution as follows:
\begin{itemize}
    \item $|0:0\rangle=|0,\cdots 0:0,\cdots 0\rangle$: Ground state.
    \item $|i:0\rangle$: $i$-th detector in group A is excited.
    \item $|0:i\rangle$: $i$-th detector in group B is excited.
    \item $|i_1:i_2\rangle$: $i_1$-th detector in group A and
    $i_2$-th detector in group B are excited.
    \item $|i_1i_2:0\rangle$: $i_1$-th and $i_2$-th detectors
    $(i_1\neq i_2)$ in group A are excited.
    \item $|0:i_1i_2\rangle$: $i_1$-th and $i_2$-th detectors
    $(i_1\neq i_2)$ in group B are excited.
    \item $C^{(0)}\equiv\left.C^{(i_1i_2)}\right|_{i_1=i_2}$,\quad
    $C^{(r)}\equiv\left.C^{(i_1i_2)}\right|_{i_1\in A,i_2\in B}$.
    \item $C^{(d)}\equiv\left.C^{(i_1i_2)}\right|_{i_1\neq i_2\in
      A}=\left.C^{(i_1i_2)}\right|_{i_1\neq i_2\in B}$. 
\end{itemize}
As we will see, the following coefficients ~\cite{Kukita2017} in the
master equation \eqref{eq:master} are necessary to calculate the
negativity,
\begin{align}
  &C_{-+}^{(r)}=\frac{2g^2e^{-\omega^2\sigma^2}}{\sqrt{\pi}}\int_{-\infty}^{+\infty}dy 
  e^{-\frac{1}{\sigma^2}(y+i\omega\sigma^2)^2}D(r_c,t+\sigma,y),\\
  &C_{++}^{(r)}=\frac{2g^2e^{-\omega^2\sigma^2}}{\sqrt{\pi}}\int_{0}^{+\infty}dy
  e^{-\frac{y^2}{\sigma^2}}D(r_c,t+\sigma+i\omega\sigma^2,y),
\end{align}
where the time coarse graining parameter $\sigma$ must satisfy
$H\sigma<1$ to guarantee the assumption to derive the master equation
\eqref{eq:master}.  The parameter $\sigma$ corresponds to the width of
switching function in analysis of the standard particle detector
model.  By applying the saddle point approximation, which is correct
for parameters with $1/\omega<\sigma<1/H, \sigma<r$, these
coefficients can be evaluated as
\begin{equation}
    C_{-+}^{(r)}=2g^2e^{-\omega^2\sigma^2}\sigma
    D(r_c,t+\sigma,-i\omega\sigma^2),
    \quad C_{++}^{(r)}=2g^2e^{-\omega^2\sigma^2}e^{2i\omega\sigma}\sigma
    D(r_c,t+\sigma+i\omega\sigma^2,0).
    \label{eq:c-conf}
\end{equation}
For the massless conformal scalar field in de Sitter spacetime with
a spatially flat time slice, these
coefficients are given by~\cite{Nambu2013,Kukita2017}
\begin{align}
 & |C_{++}^{(r)}|=\frac{g ^2 \sigma H^2
 e^{-\omega^2\sigma^2}}{2 \pi 
   ^2}\frac{1}{H^2r^2},\quad
  C_{-+}^{(r)}=\frac{g ^2 \sigma H^2  e^{-\omega^2\sigma^2}}{2 \pi ^2}\frac{1}{
H^2r^2 +4 \sin
   ^2\theta},
\label{eq:c-mini}
\end{align}
where $\theta\equiv H\sigma^2\omega<\pi$,  and
$r=e^{N}r_c\le e^{N} H^{-1}$ denotes the physical separation between
detectors at $e$-folding time of inflation $N=H\times(t_0+\sigma)$.  For the
massless minimal scalar field,
\begin{align}
    &|C_{++}^{(r)}|=\frac{g^2\sigma H^2e^{-\omega^2\sigma^2}}{2\pi^2}
\left|\frac{e^{-2i\theta}}{H^2r^2}+1-\frac{1}{2}
\left\{\mathrm{Ei}(ik_0e^{-N}r)+\mathrm{Ei}(-ik_0e^{-N}r)
\right\}\right|,
      \notag \\
  &C_{-+}^{(r)}=\frac{g^2\sigma H^2e^{-\omega^2\sigma^2}}{2\pi^2}\Biggl[
    \frac{1}{H^2r^2+4\sin^2\theta} \notag \\
&\qquad
  -\frac{1}{2}\left\{
    \mathrm{Ei}(-ik_0e^{-N}(r-2iH^{-1}\sin\theta))+
    \mathrm{Ei}(-ik_0e^{-N}(-r
-2iH^{-1}\sin\theta))\right\}+1\Biggr],
\end{align}
where $k_0$ is the infrared cutoff corresponding to comoving size of
the inflating universe $k_0=H$ and
 $\mathrm{Ei}(-x)=-\int_x^\infty\frac{dy}{y}e^{-y}$ is the
exponential integral.

 As a warming up, we first review
$1+1$ detectors case, which is often adopted as a model of entanglement
harvesting in numerous situations.

\section{Negativity for $1+1$ detectors system (two qubits case) }
For the initial separable state of detectors (We adopt the
  basis $\{|1:1\rangle,|1:0\rangle,|0:1\rangle,|0:0\rangle \}$ which is
  descending order of states in binary numbering.)
\begin{equation}
  \rho_0=|0:0\rangle\langle 0:0|=
  \begin{pmatrix} 0 & 0 & 0 & 0\\
    0 & 0 & 0 & 0\\
    0 & 0 & 0 & 0\\
    0 & 0 & 0 & 1
  \end{pmatrix},
\end{equation}
the state of detectors \eqref{eq:sol} becomes
\begin{equation}
  \rho=\begin{pmatrix}
    0 & 0 & 0 & -C_{++}^{(r)}\Delta t \\
    0 & C_{-+}^{(0)}\Delta t & C_{-+}^{(r)}\Delta t& 0 \\
    0 & C_{-+}^{(r)}\Delta t &C_{-+}^{(0)}\Delta t & 0 \\
    -C_{--}^{(r)}\Delta t & 0 & 0 & 1-2C_{-+}^{(0)}\Delta t
    \end{pmatrix}.
\label{eq:state}
\end{equation}
To quantify entanglement between detectors, we introduce the
entanglement negativity~\cite{Peres1996,Horodecki1996}
\begin{equation}
    E_N=\sum_{\lambda_i<0}|\lambda_i|,
\end{equation}
where $\lambda_i$ are eigenvalues of partially transposed state
$\rho^\text{PT}$, which is defined by transposing components belonging
to group B only.
For the state \eqref{eq:state},
\begin{equation}
  \rho^\text{PT}=\begin{pmatrix}
    0 & 0 & 0 & C_{-+}^{(r)}\Delta t \\
    0 & C_{-+}^{(0)}\Delta t & -C_{++}^{(r)}\Delta t& 0 \\
    0 & -C_{--}^{(r)}\Delta t &C_{-+}^{(0)}\Delta t & 0 \\
    C_{-+}^{(r)}\Delta t & 0 & 0 & 1-2C_{-+}^{(0)}\Delta t
    \end{pmatrix}.
\label{eq:state2}
\end{equation}
Eigenvalues of this state are
\begin{equation}
    \lambda=1-2\Delta tC_{-+}^{(0)},~\Delta t\left( C_{-+}^{(0)}\pm
        \left|C_{++}^{(r)}\right|
    \right).
\end{equation}
As we are considering $\omega\Delta t\ll 1$ with the weak coupling
limit, which implies the coefficient $|C|=O(g^2)\ll 1$. Hence, the only
eigenvalue which can become negative is
$\Delta t\left(C_{-+}^{(0)}-\left|C_{++}^{(r)}\right|\right)$ and the
negativity is given by
$E_N=\Delta
t\,\text{max}\left[\left(|C_{++}^{(r)}|-C_{-+}^{(0)}\right),0\right]$.
As the sign of negativity does not depends on $\Delta t$, the initial
separable state can become entangled one instantly.  From now on in
this paper, we designate the following quantity as the negativity
\begin{equation}
  E_N=|C_{++}^{(r)}|-C_{-+}^{(0)}.
  \label{eq:negativity}
\end{equation}
By this definition of the negativity, the state is entangled for
$E_N>0$. For two qubits case ($m=n=1$), $E_N<0$ implies the state is
separable.  However, for $m+n\ge3$, we can say nothing about
separability from the condition $E_N<0$. Comparing the formula of the
negativity \eqref{eq:negativity} with that derived
previously~\cite{Nambu2013} using a switching function of detectors,
Equation \eqref{eq:negativity} exactly coincides with previous one and we can
confirm that the time coarse graining parameter $\sigma$ has a meaning
of a width of switching function of detectors.

\begin{figure}[H] 
  \centering
  \includegraphics[width=0.38\linewidth,clip]{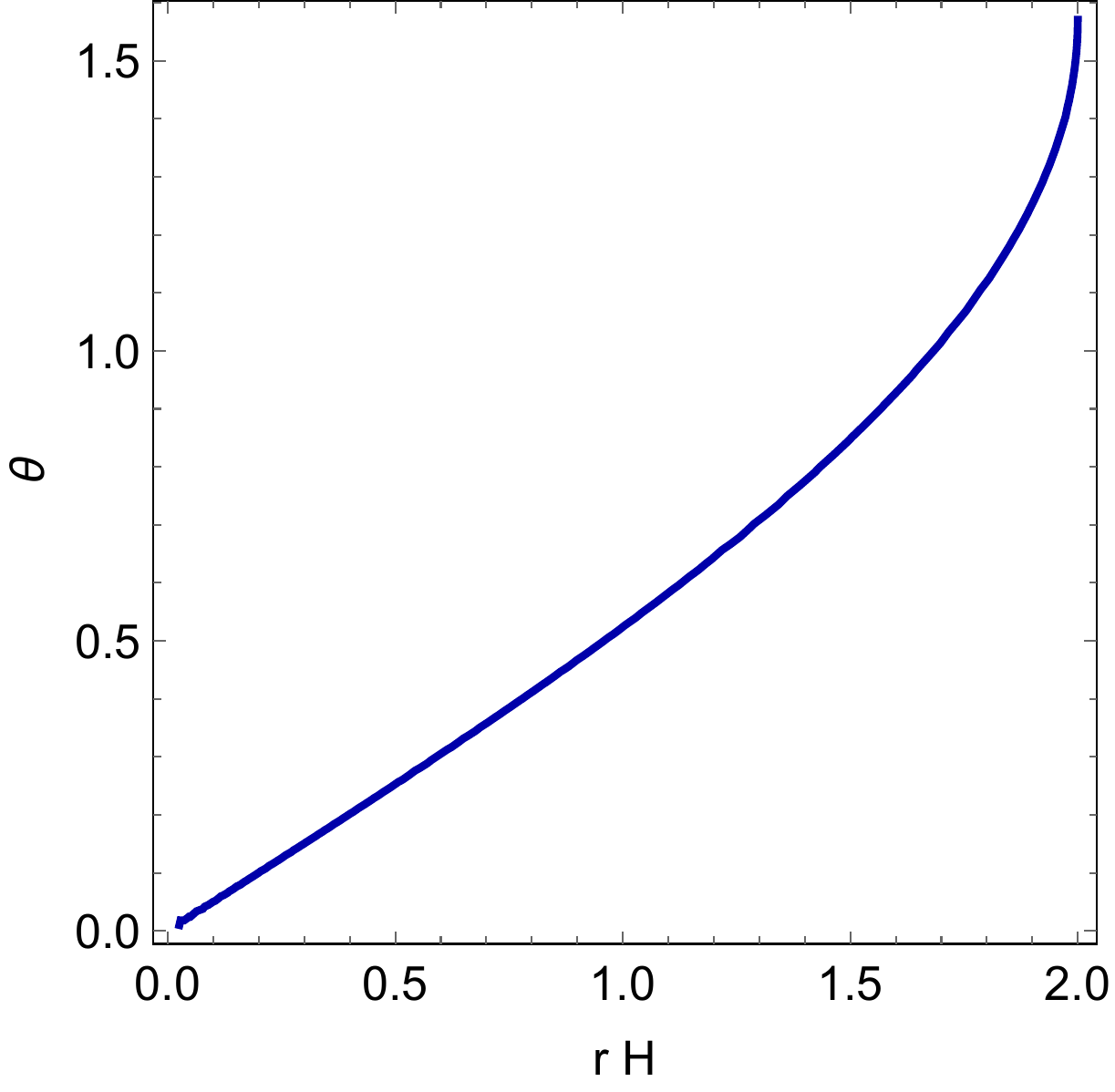}%
  \hspace{2.0cm}
  \includegraphics[width=0.38\linewidth,clip]{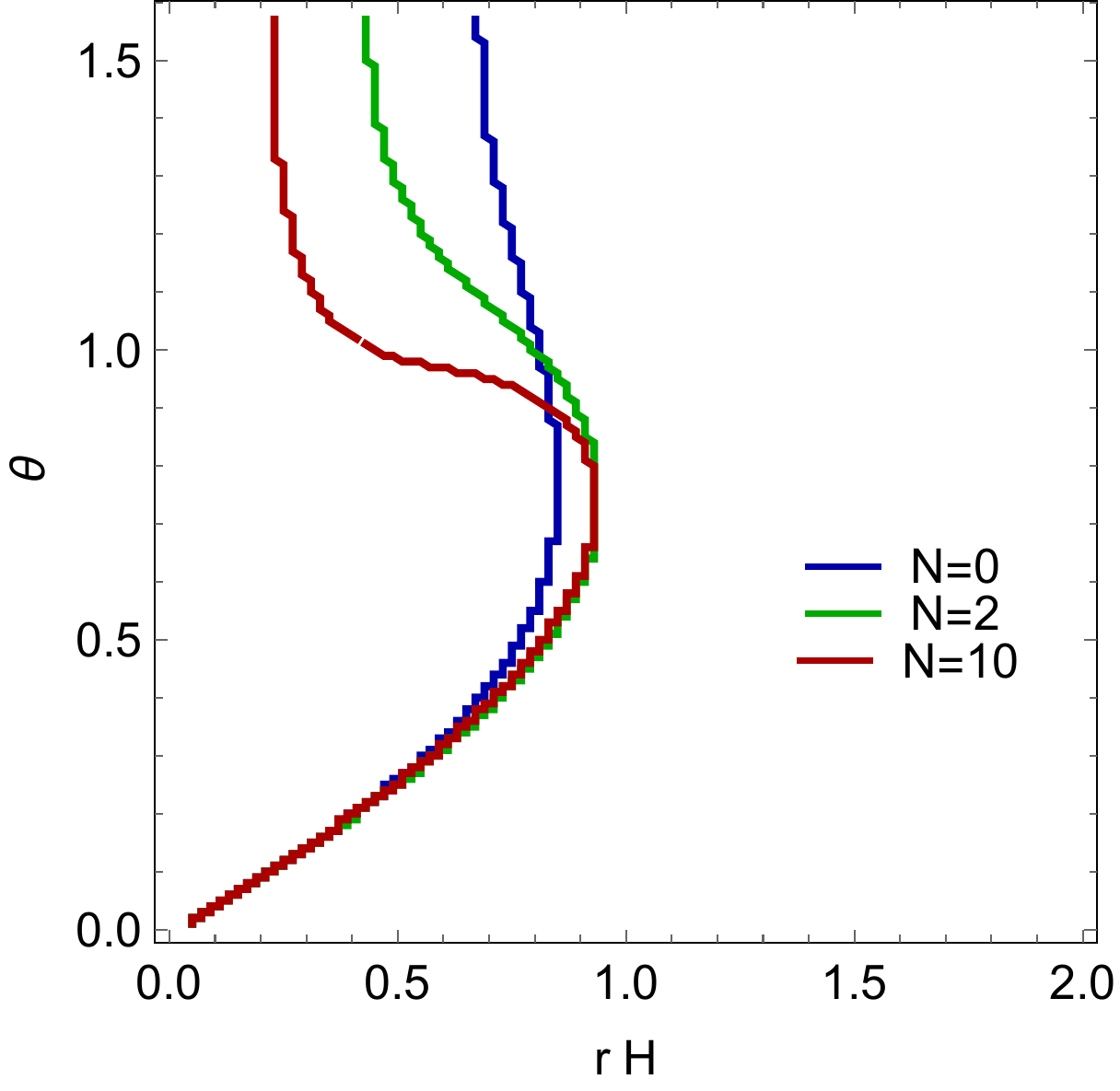}
  \caption{The entanglement detection is possible for parameters
    $(r,\theta)$ in  regions left side of each solid lines. Left panel:
    massless conformal scalar field. Right panel: massless minimal
    scalar field with $e$-foldings $N=0,2,10$.}
  \label{fig:region}
\end{figure}

Fig.~\ref{fig:region} shows parameter regions where entanglement
detection is possible. For parameters $(r, \theta)$ belonging to left
side regions of each lines, the negativity is positive and a pair of
detectors can harvest the entanglement of the scalar field.
\noindent
For the massless conformal scalar field, the maximum distance of
entanglement detection is $2H^{-1}$ and for the massless minimal
scalar field, that distance is $H^{-1}$.  In this paper, we regard the
Hubble length $H^{-1}$ as the horizon radius and define the super
horizon scale as $r>2H^{-1}$. Following this definition of ``super
horizon scale'', a pair of detectors cannot access to entanglement on
the super horizon scale for both type of scalar fields.

\section{Negativity for $m+n$ detectors system}
Let us consider entanglement harvesting using $m+n$ detectors.
The initial state of detectors is assumed to be
\begin{equation}
    \rho_0=|0:0\rangle\langle 0:0|,
\end{equation}
and the state after time evolution is obtained by evaluating
$\mathcal{L}[\rho_0]$ in Equation \eqref{eq:sol}:
\begin{align}
    \rho&=\rho_0+\frac{\Delta t}{2}\sum_{\al_1,\al_2=1}^{m+n}\sum_{j_1,j_2=\pm}C_{j_1j_2}^{(\al_1\al_2)}\left(2\sigma_{j_2}^{(\al_2)}\rho_0\,\sigma_{j_1}^{(\al_1)}
        -\sigma_{j_1}^{(\al_1)}\sigma_{j_2}^{(\al_2)}\rho_0
 -\rho_0\,\sigma_{j_1}^{(\al_1)}\sigma_{j_2}^{(\al_2)}\right) \notag \\
  &=\rho_0 \notag \\
    &+\Delta t\Biggl[\sum_{i_1,i_2\in
      A}\!\!C_{-+}^{(i_1i_2)}|i_1:0\rangle\langle
      i_2:0|+\sum_{\substack{i_1\in A \\
      i_2\in B}}C_{-+}^{(i_1i_2)}|i_1:0\rangle\langle0:i_2| \notag \\
    &\qquad\qquad\qquad\qquad\qquad
      +\sum_{\substack{i_1\in B \\ i_2\in A}}C_{-+}^{(i_1i_2)}|0:i_1\rangle\langle
      i_2:0|+\sum_{i_1,i_2\in
      B}\!\!C_{-+}^{(i_1i_2)}|0:i_1\rangle\langle0:i_2|
      \Biggr] \notag\\
  &-\frac{\Delta t}{2}\Biggl[2\sum_{\substack{i_1\in A \\ i_2\in
    B}}C_{++}^{(i_1i_2)}|i_1:i_2\rangle\langle
    0:0|+\sum_{i_1\neq i_2\in
    A}\!\!C_{++}^{(i_1i_2)}|i_1i_2:0\rangle\langle 0:0| \notag \\
  &\qquad\qquad\qquad\qquad\qquad
    +\sum_{i_1\neq
    i_2\in B}\!\!C_{++}^{(i_1i_2)}|0:i_1i_2\rangle\langle 0:0|+(m+n)
    C_{-+}^{(0)}|0:0\rangle\langle 0:0| \Biggr] \notag\\
&-\frac{\Delta t}{2}\Biggl[2\sum_{\substack{i_1\in A \\i_2\in
    B}}C_{--}^{(i_1i_2)}|0:0\rangle\langle i_1:i_2|
  +\sum_{i_1\neq i_2\in
    A}\!\!C_{--}^{(i_1i_2)}|0:0\rangle\langle i_1i_2:0|\notag \\
  &\qquad\qquad\qquad\qquad\qquad
    +\sum_{i_1\neq
    i_2\in B}\!\!C_{--}^{(i_1i_2)}|0:0\rangle\langle 0:i_1i_2|+(m+n)
    C_{-+}^{(0)}|0:0\rangle\langle 0:0| \Biggr].
\end{align}
After partial transposition of the state with respect to the group B,
\begin{align}
    &\rho^{\text{PT}}
  =\rho_0 \notag \\
    &+\Delta t\Biggl[\sum_{i_1,i_2\in
      A}\!\!C_{-+}^{(i_1i_2)}|i_1:0\rangle\langle
      i_2:0|+\sum_{\substack{i_1\in A \\
      i_2\in B}}C_{-+}^{(i_1i_2)}|i_1:i_2\rangle\langle0:0| \notag\\
  &\qquad\qquad\qquad\qquad\qquad\qquad
      +\sum_{\substack{i_1\in B \\i_2\in A}}C_{-+}^{(i_1i_2)}|0:0\rangle\langle
      i_2:i_1|+\sum_{i_1,i_2\in
      B}\!\!C_{-+}^{(i_1i_2)}|0:i_2\rangle\langle0:i_1|
      \Biggr] \notag\\
  &-\frac{\Delta t}{2}\Biggl[2\sum_{\substack{i_1\in A\\i_2\in
    B}}C_{++}^{(i_1i_2)}|i_1:0\rangle\langle
    0:i_2|+\sum_{i_1\neq i_2\in
    A}\!\!C_{++}^{(i_1i_2)}|i_1i_2:0\rangle\langle 0:0| \notag \\
  &\qquad\qquad\qquad\qquad\qquad\qquad
    +\sum_{i_1\neq
    i_2\in B}\!\!C_{++}^{(i_1i_2)}|0:0\rangle\langle 0:i_1i_2|+(m+n)
    C_{-+}^{(0)}|0:0\rangle\langle 0:0| \Biggl] \notag\\
&-\frac{\Delta t}{2}\Biggl[2\sum_{\substack{i_1\in A \\i_2\in
    B}}C_{--}^{(i_1i_2)}|0:i_2\rangle\langle i_1:0|
  +\sum_{i_1\neq i_2\in
    A}\!\!C_{--}^{(i_1i_2)}|0:0\rangle\langle i_1i_2:0| \notag \\
  &\qquad\qquad\qquad\qquad\qquad\qquad
    +\sum_{i_1\neq
    i_2\in B}\!\!C_{--}^{(i_1i_2)}|0:i_1i_2\rangle\langle 0:0|+(m+n)
    C_{-+}^{(0)}|0:0\rangle\langle 0:0| \Biggr] \notag \\
& \notag\\
 &=\Delta t\Biggl[\sum_{i_1,i_2\in
      A}C_{-+}^{(i_1i_2)}|i_1:0\rangle\langle i_2:0|      
      +\sum_{i_1,i_2\in
      B}C_{-+}^{(i_1i_2)}|0:i_2\rangle\langle0:i_1|    \notag \\
  &\qquad\qquad\qquad\qquad\qquad\qquad
    -\sum_{\substack{i_1\in A \\i_2\in
    B}}C_{++}^{(i_1i_2)}|i_1:0\rangle\langle
    0:i_2|
-\sum_{\substack{i_1\in A \\i_2\in
    B}}C_{--}^{(i_1i_2)}|0:i_2\rangle\langle i_1:0|
   \Biggr] \notag \\
 &\quad +\left(1-\Delta t(m+n)
C_{-+}^{(0)}\right)|0:0\rangle\langle 0:0| \notag \\
 &\qquad\qquad\qquad\qquad\qquad
   +\Delta
   t\Biggl[\sum_{\substack{i_1\in A \\i_2\in
  B}}C_{-+}^{(i_1i_2)}|i_1:i_2\rangle\langle0:0| 
      +\sum_{\substack{i_1\in B \\i_2\in A}}C_{-+}^{(i_1i_2)}|0:0\rangle\langle
      i_2:i_1| \Biggr] \notag \\
 &\quad  -\frac{\Delta t}{2} \left[\sum_{i_1\neq i_2\in
    A}\!\!C_{++}^{(i_1i_2)}|i_1i_2:0\rangle\langle 0:0|
    +\sum_{i_1\neq
    i_2\in B}\!\!C_{++}^{(i_1i_2)}|0:0\rangle\langle 0:i_1i_2|\right]
       \notag \\
&\quad -\frac{\Delta t}{2}\left[\sum_{i_1\neq i_2\in
    A}\!\!C_{--}^{(i_1i_2)}|0:0\rangle\langle i_1i_2:0|
  +\sum_{i_1\neq
    i_2\in B}\!\!C_{--}^{(i_1i_2)}|0:i_1i_2\rangle\langle 0:0| \right] \notag \\
& \notag \\
&\equiv \rho_1+\rho_2,
\end{align}
where $\rho_1$ denotes a part of $\rho^{\text{PT}}$ spanned by the
basis $\{|i_1\!:\!0\rangle, |0\!:\!i_1\rangle\}$ and $\rho_2$ is a part of
$\rho^{\text{PT}}$ spanned by the basis
$\{|0\!:\!0\rangle, |i_1\!:\!i_2\rangle, |i_1i_2\!:\!0\rangle,
|0\!:\!i_1i_2\rangle\}$.
As these two sets of basis are orthogonal to each other, a matrix
representation of $\rho^{\text{PT}}$ with these basis has a block
diagonal structure. Hence to obtain eigenvalues of $\rho^{\text{PT}}$,
only we have to do is to consider eigenvalues of $\rho_1$ and $\rho_2$
separately.

\subsection{Eigenvalues of $\rho_1$}
The eigenvalue equation is
\begin{equation}
    \rho_1|\lambda\rangle=\lambda|\lambda\rangle.
\end{equation}
From the configuration of detectors we are considering, we can assume
the following form of the eigenvector
\begin{equation}
    |\lambda\rangle=\al\sum_{i\in A}|i:0\rangle+\beta\sum_{i\in B}|0:i\rangle,
\end{equation}
where $\al,\beta$ are coefficients to be determined. By applying $\rho_1$,
\begin{align}
  \rho_1|\lambda\rangle&=\al\sum_{i_1,i_2\in
                         A}C_{-+}^{(i_1i_2)}|i_1:0\rangle
                         -\al\sum_{\substack{i_1\in A \\ i_2\in
                         B}}C_{--}^{(i_1i_2)}|0:i_2\rangle+\beta
          \sum_{i_1,i_2\in
                         B}\!\!C_{-+}^{(i_1i_2)}|0:i_2\rangle
-\beta\sum_{\substack{i_1\in A \\i_2\in B}}C_{++}^{(i_1i_2)}|i_1:0\rangle
                         \notag \\
  &=\sum_{i_1\in A}\left(\al\sum_{i_2\in A}C_{-+}^{(i_1i_2)}
    -\beta\sum_{i_2\in
    B}C_{++}^{(i_1i_2)}\right)|i_1:0\rangle+\sum_{i_1\in B}\left(
   -\al\sum_{i_2\in A}C_{--}^{(i_1i_2)}+\beta\sum_{i_2\in
    B}C_{-+}^{(i_1i_2)} \right)|0:i_1\rangle,
\end{align}
and the eigenvalue equation is reduced to be
\begin{align}
  &\lambda\, \al=\al\sum_{i_2\in A}C_{-+}^{(i_1i_2)}-\beta\sum_{i_2\in
    B}C_{++}^{(i_1i_2)}\quad\text{with}~ i_1\in A,\\
&\lambda\, \beta=-\al\sum_{i_2\in A}C_{--}^{(i_1i_2)}+\beta\sum_{i_2\in
    B}C_{-+}^{(i_1i_2)}\quad\text{with}~ i_1\in B.
\end{align}
Thus,
\begin{equation}
  \lambda\al=\al C_{-+}^{(0)}+\al(m-1)C_{-+}^{(d)}-\beta n
    C_{++}^{(r)},\quad
    \lambda\beta=\beta C_{-+}^{(0)}+\beta(n-1)C_{-+}^{(d)}-\al m
      C_{--}^{(r)}.
\end{equation}
By eliminating $\al,\beta$, we obtain
\begin{equation}
    \left(\lambda-C_{-+}^{(0)}-(m-1)C_{-+}^{(d)}\right)
 \left(\lambda-C_{-+}^{(0)}-(n-1)C_{-+}^{(d)}\right)
    =mn\left|C_{++}^{(r)}\right|^2,
\end{equation}
and eigenvalues are
\begin{equation}
    \lambda=\left[
    C_{-+}^{(0)}+\left(\frac{m+n}{2}-1\right)C_{-+}^{(d)}\right]\pm
\left[mn\left|C_{++}^{(r)}\right|^2+\left(\frac{m-n}{2}\right)^2
\left(  C_{-+}^{(d)}\right)^2\right]^{1/2}.
\label{eq:eigen1}
\end{equation}
These quantities are $O(g^2)$.
\subsection{Eigenvalues of $\rho_2$}
We will show that eigenvalues of $\rho_2$ are positive up to $O(g^2)$
and they do not contribute to the negativity in the present order of
calculation. We assume  the form of the eigenvector as
\begin{equation}
    |\lambda\rangle=|0:0\rangle+\sum_{\substack{i_1\in A \\i_2\in
      B}}\al_{i_1i_2}|i_1:i_2\rangle+\!\!\sum_{i_1\neq i_2\in
      A}\!\!\beta_{i_1i_2}|i_1i_2:0\rangle +\!\!\sum_{i_1\neq i_2\in
      B}\!\!\gamma_{i_1i_2}|0:i_1i_2\rangle,
\end{equation}
where $\al,\beta,\gamma$ are coefficients to be determined.  By
applying $\rho_2$,
\begin{align}
  &\rho_2|\lambda\rangle=\left(1-\Delta t(m+n)C_{-+}^{(0)}\right)|0:0\rangle
                         \notag \\
&\quad +\frac{\Delta t}{2}\Biggl[2\sum_{\substack{i_1\in A \\i_2\in
  B}}C_{-+}^{(i_1i_2)}|i_1:i_2\rangle-\sum_{i_1\neq i_2\in
  A}\!\!C_{++}^{(i_1i_2)}|i_1i_2:0\rangle-\sum_{i_1\neq i_2\in
  B}\!\!C_{--}^{(i_1i_2)}|0:i_1i_2\rangle \Biggr] \notag \\
&\quad +\Delta t\Biggl[\sum_{\substack{i_1\in A \\i_2\in
  B}}\al_{i_1i_2}C_{-+}^{(i_1i_2)}|0:0\rangle  \Biggr] 
-\frac{\Delta t}{2}\Biggl[\sum_{i_1\neq i_2\in
  A}\!\!\!\beta_{i_1i_2}C_{--}^{(i_1i_2)}|0:0\rangle
  +\sum_{i_1\neq i_2\in B}\!\!\!\gamma_{i_1i_2}C_{++}^{(i_1i_2)}|0:0\rangle\Biggr].
\end{align}
From this, we have the following equations
\begin{align}
  &\lambda=\left(1-\Delta t(m+n)C_{--}^{(0)}\right)
+\frac{\Delta t}{2}\Biggl(2\sum_{\substack{i_1\in A \\i_2\in
    B}}\al_{i_1i_2}C_{-+}^{(i_1i_2)}-\!\!\sum_{i_1\neq i_2\in
    A}\!\!\beta_{i_1i_2}C_{--}^{(i_1i_2)}-\!\!\sum_{i_1\neq i_2\in
    B}\!\!\gamma_{i_1i_2}C_{++}^{(i_1i_2)}\Biggr), \\
  &\lambda\al_{i_1i_2}=\Delta t C_{-+}^{(i_1i_2)}\quad (i_1\in A, i_2\in
    B),\\
  &\lambda\beta_{i_1i_2}=-\frac{\Delta t}{2} C_{++}^{(i_1i_2)}\quad
    (i_1\neq i_2\in A), \\
  &\lambda\gamma_{i_1i_2}=-\frac{\Delta t}{2} C_{--}^{(i_1i_2)}\quad (i_1\neq i_2\in B).
\end{align}
After eliminating coefficients $\al, \beta, \gamma$, we obtain
\begin{equation}
    \lambda^2-\left(1-\Delta t(m+n)C_{-+}^{(0)}\right)\lambda-(\Delta t)^2\Biggl[
        \sum_{\substack{i_1\in A \\i_2\in
          B}}\left(C_{-+}^{(i_1i_2)}\right)^2+\frac{1}{4}
        \sum_{i_1\neq
          i_2\in A}\!\!\left|C_{++}^{(i_1i_2)}\right|^2+
        \frac{1}{4}\sum_{i_1\neq
          i_2\in B}\!\!\left|C_{++}^{(i_1i_2)}\right|^2\Biggr]=0,
\end{equation}
and eigenvalues are
\begin{align}
    \lambda&=\frac{1}{2}\left(1-\Delta t(m+n)C_{-+}^{(0)}\right)\pm\frac{1}{2}
    \left[\left(1-\Delta t(m+n)C_{-+}^{(0)}
        \right)^2+\text{terms}\left|C\right|^2 \right]^{1/2} \notag \\
  &=    1-\Delta t(m+n)C_{-+}^{(0)}+O(g^4),\quad  O(g^4).
\label{eq:eigen2}
\end{align}
Thus,  up to $O(g^2)$, $\rho_2$ does not have negative eigenvalues.

\subsection{Negativity}
\begin{figure}[H]
    \centering
    \includegraphics[width=0.5\linewidth,clip]{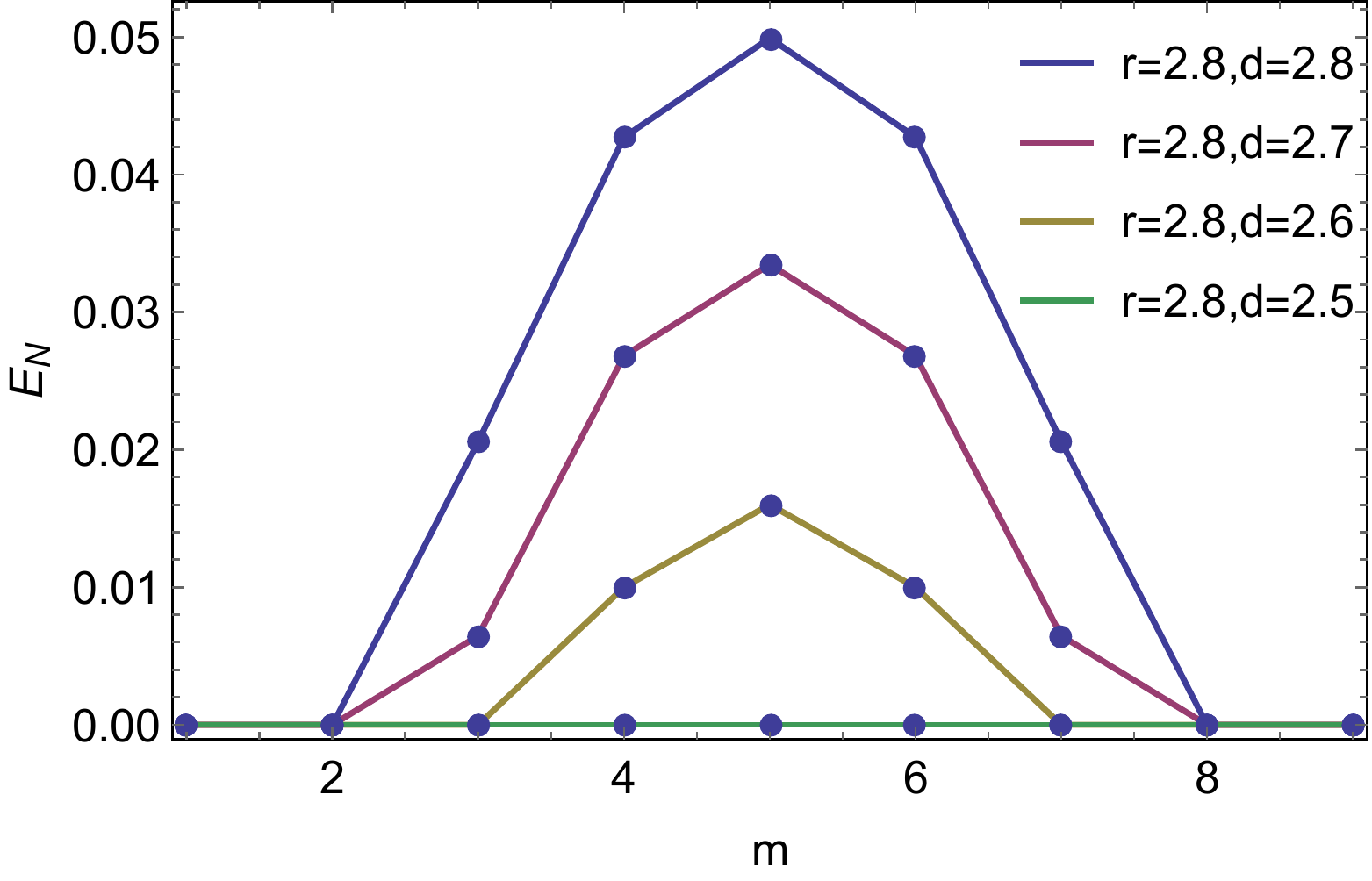}
    \caption{Dependence of number $m$ of group A on negativity (the
      massless conformal scalar field with $\theta=\pi/2$).  The total
      number of detectors is $m+n=10$ with $r=2.8H^{-1}$ (super
      horizon scale). The maximum value of the negativity is attained
      for $m=n=5$. }
\label{fig:bi}
\end{figure}

\begin{figure}[H]
  \centering
   \includegraphics[width=0.38\linewidth,clip]{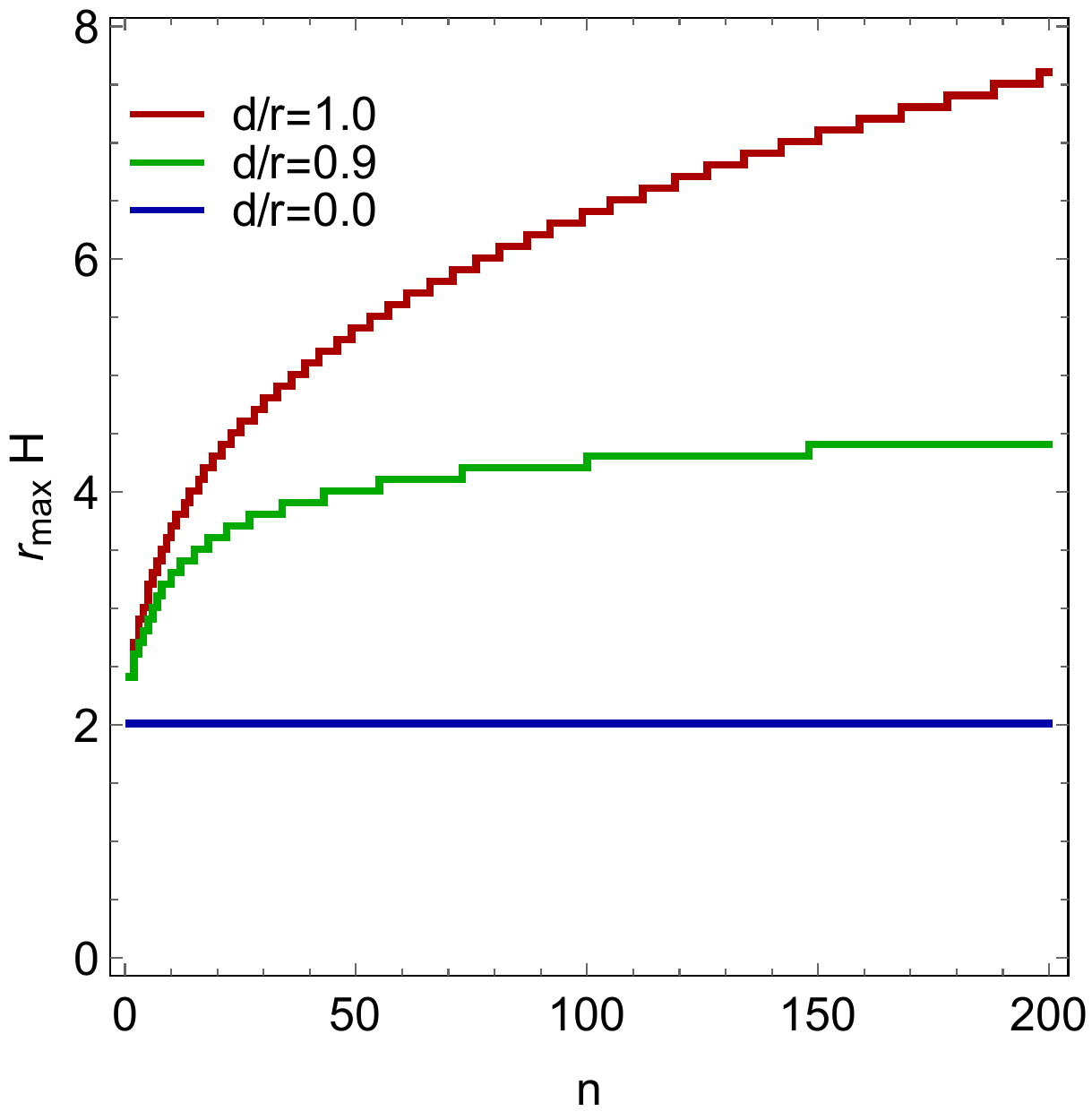}%
   \hspace{2cm}
   \includegraphics[width=0.38\linewidth,clip]{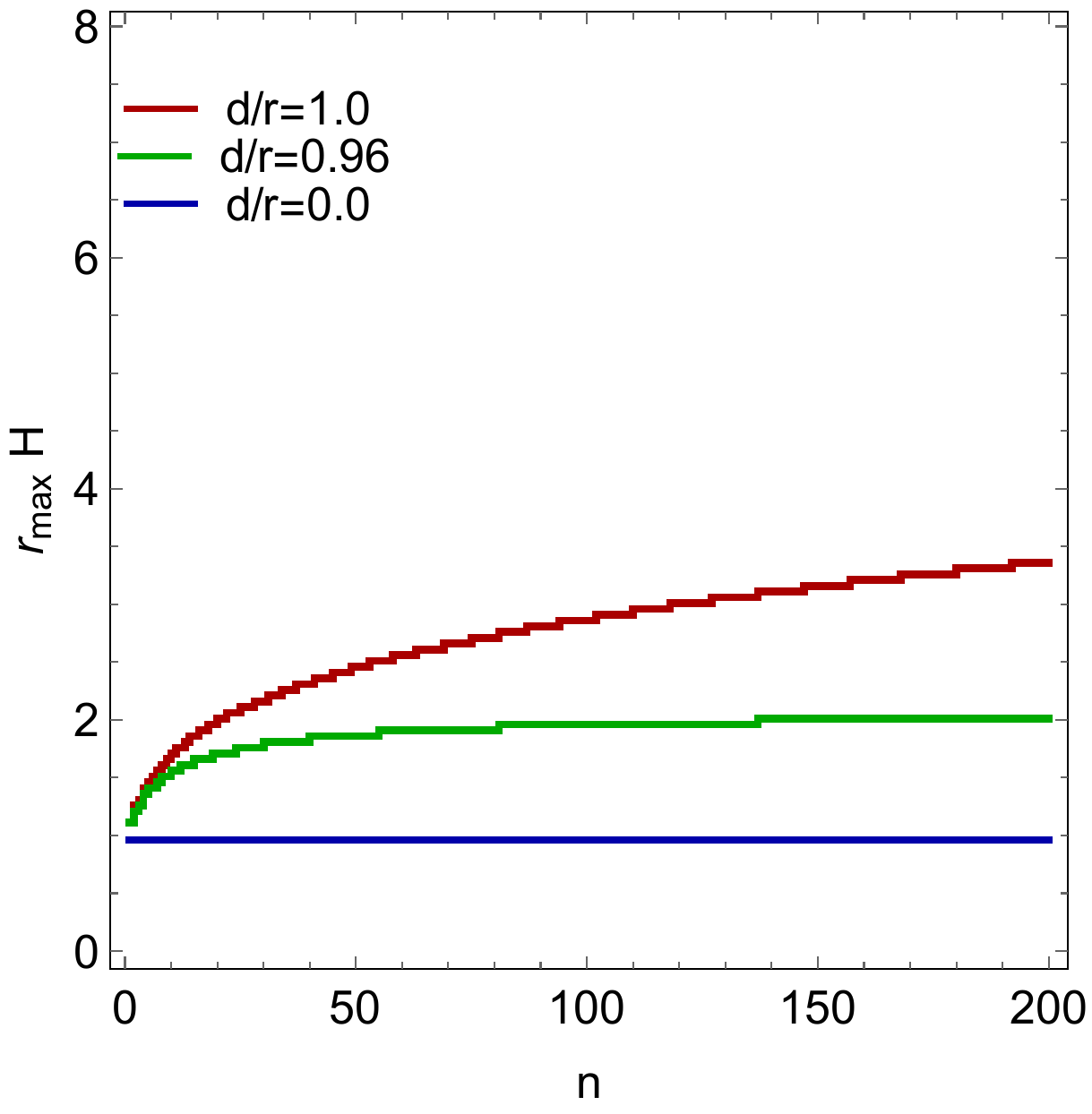}
   \caption{ $r_\text{max}$ as a function of number of detectors $n$.
     Left panel: the massless conformal scalar field with
     $\theta=\pi/2$. Right panel: the massless minimal scalar field
     with $\theta=\pi/4, N=10$. For both types of scalar fields, with $d/r=1$,
     $r_\text{max}$ exceeds the horizon scale $2H^{-1}$ as the number
     of detectors increases.} 
 \label{fig:rmax}
\end{figure}

As the $\rho_2$ does not have negative eigenvalues, the negativity of
the bipartite state $\rho$ is given by the eigenvalue of $\rho_1$ and
we obtain the following key formula of the negativity in this paper
\begin{equation}
    E_N=\left[mn\left|C_{++}^{(r)}\right|^2+\left(\frac{m-n}{2}\right)^2\left(
            C_{-+}^{(d)}\right)^2\right]^{1/2}-\left[
    C_{-+}^{(0)}+\left(\frac{m+n}{2}-1\right)C_{-+}^{(d)}\right].
\end{equation}
For a fixed value of the total number of detectors $m+n$,
\begin{equation}
  \frac{\pa E_N}{\pa m}=\frac{n-m}{2}\left[mn\left|C_{++}^{(r)}\right|^2+\left(\frac{m-n}{2}\right)^2\left(
            C_{-+}^{(d)}\right)^2\right]^{-1/2}\left[\left|C_{++}^{(r)}\right|^2+
\left(            C_{-+}^{(d)}\right)^2\right],
\end{equation}
and if the total number of detectors is even, $m=n$ provides a maximum
value of the negativity
\begin{equation}
    E_N=n\left(\left|C_{++}^{(r)}\right|
-\frac{C_{-+}^{(0)}+(n-1)C_{-+}^{(d)}}{n}\right).
\end{equation}
Fig.~\ref{fig:bi} shows an example of detection of super horizon
scale entanglement with ten detectors. In this case, ten
detectors catch nonzero negativity on the super horizon scale  $r=2.8H^{-1}$.
\noindent
Using Equation \eqref{eq:c-conf}, the negativity for the massless
conformal scalar field is
\begin{equation}
  E_N\propto
n\left[\frac{1}{H^2r^2}-\frac{1}{n}
    \left(\frac{1}{4\sin^2\theta}
+\frac{n-1}{4\sin^2\theta+H^2d^2}\right)\right].
\end{equation}
For a given $n$ and separation $d$ of detectors in the same group, we
introduce  a ratio $\del=d/r$ with $0\le\del\le 1$. Then the maximum
distance of entanglement detection is given by
\begin{equation}
  r_{\text{max}}=\frac{\sqrt{2}\sin\theta}{H}\left(\frac{1-\del^2}{\del^2}
\right)^{1/2}n^{1/4}
\left[\left(n+\frac{4\del^2}{(1-\del^2)^2}\right)^{1/2}-n^{1/2}\right]^{1/2}.
\end{equation}
For $\del=1$, we have
\begin{equation}
  r_\text{max}=2H^{-1}n^{1/4}\sin\theta,
\end{equation}
and $r_\text{max}$ can become super horizon scale as $n$
increases. For $\del<1$, in the limit of large $n\gg1$, $r_\text{max}$
approaches the following asymptotic value
\begin{equation}
  r_\text{max}\sim  \dfrac{2H^{-1}\sin\theta}{\sqrt{1-\del^2}}.
\end{equation}
For the massless minimal scalar field, although we cannot obtain the
analytic expression of $r_\text{max}$, it is possible to obtain the
 asymptotic formula. For $n\gg1$ with $\del\approx 1$,
\begin{equation}
  r_\text{max}\sim
  \begin{cases}
    (2\sin 2\theta)^{1/2}\left(\dfrac{n}{\ln n}\right)^{1/4}H^{-1}\quad
    &\text{for}\quad\del=1, \\
 \dfrac{(\sin2\theta)^{1/2}}{\del(-\ln\del)^{1/4}}H^{-1}\quad&\text{for}\quad
    \del\neq1.
  \end{cases}
\end{equation}
For $\del\neq 1$ with $\theta=\pi/4$, $r_\text{max}$ can exceeds
$2H^{-1}$ for $\del$ greater than $\approx 0.96$.
Fig.~\ref{fig:rmax} shows $r_\text{max}$ as a function of number of
detectors $n$.
\noindent
For both type of scalar fields, $r_\text{max}$ approaches constant
values as $n\rightarrow\infty$ for $\del=d/r<1$. These constant values
become larger than $2H^{-1}$ if $\del$ is sufficiently close to
unity. If we take $\del=1$,
$r_\text{max}$ grows as the number of detectors increases and
 becomes infinity as $n\rightarrow\infty$.  Therefore, it is possible to detect
the super horizon scale entanglement if we prepare sufficiently large
number of detectors.

\section{Monogamy inequality}
We expect that detectability of large scale entanglement on the super
horizon scale is related to multipartite entanglement. To make the
connection clear, we check the monogamy inequality of
negativity~\cite{Ou2007} for the present $m+n$ detectors system. For a
tripartite system $A\cup B_1\cup B_2$, the negativity between $A$ and
$B_1$, between $A$ and $B_2$ and between $A$ and $B_1B_2$ should obey
the following monogamy inequality
\begin{equation}
  E_N^2(A:B_1)+E_N^2(A:B_2)\le E_N^2(A:B_1B_2).
\end{equation}
This inequality implies
\begin{equation}
  \sum_{j_1,j_2}(E_N(A_{j_1}:B_{j_2}))^2\le (E_N(A:B))^2,
  \label{eq:mono}
\end{equation}
where $A=\cup_{j=1,\cdots,m} A_j$ and $B=\cup_{j=1,\cdots,n} B_j$.
For the present detectors model with $m=n$, the left hand side of
the inequality \eqref{eq:mono} is
\begin{equation}
  n^2\times\left(|C_{++}^{(r)}|-C_{-+}^{(0)}\right)^2,
\end{equation}
and the right hand side of the inequality \eqref{eq:mono} is
\begin{equation}
  n^2\left(|C_{++}^{(r)}|-\frac{C_{-+}^{(0)}+(n-1)C_{-+}^{(d)}}{n}\right)^2=
  n^2\left[\left|C_{++}^{(r)}\right|-C_{-+}^{(0)}+\left(1-\frac{1}{n}\right)
  \left(C_{-+}^{(0)}-C_{-+}^{(d)}\right)\right]^2.
\end{equation}
As $C_{-+}^{(0)}\ge C_{-+}^{(d)}$ which can be directly confirmed from
Equations \eqref{eq:c-conf} and \eqref{eq:c-mini}, the inequality
\eqref{eq:mono} definitely holds.

The difference between both side of the inequality \eqref{eq:mono} can
be interpreted as the residual entanglement and regarded as
quantifying degrees of multipartite entanglement~\cite{Ou2007}. This
quantity is
\begin{equation}
  n^2\left[\left(1-\frac{1}{n}\right)^2\left(C_{-+}^{(0)}-C_{-+}^{(d)}\right)^2+2
\left(1-\frac{1}{n}\right)\left(\left|C_{++}^{(r)}\right|-C_{-+}^{(0)}\right)
\left(C_{-+}^{(0)}-C_{-+}^{(d)}\right) \right].  
\end{equation}
If we take $d=0$, this difference becomes zero and $r_\text{max}$ 
reduces to  $2H^{-1}$ for the massless conformal
scalar case, which is the same value attained by a pair of detectors. The
residual entanglement becomes maximum for $d=r$, in which case $r_\text{max}$
can become larger than the Hubble horizon scale provided that
sufficiently large number of detectors are prepared. Thus, effect of
multipartite entanglement is crucial for detection of the bipartite
entanglement on the super horizon scale.


\section{Summary}

We investigated detection of entanglement of the scalar field on super
horizon scale in de Sitter space using multiple detectors. For this
purpose, we obtained the formula of negativity for $m+n $ qubit
detectors system.  The maximum possible distance of detecting nonzero values of
negativity is bounded by $n^{1/4}H^{-1}$ for the massless conformal
scalar field and $(n/\ln n)^{1/4}H^{-1}$ for the massless minimal
scalar field. For both type of scalar fields, these bounds grow as the
number of detectors increases. Thus, it is possible to detect
entanglement on the super horizon scale if we prepare sufficient large
number of detectors.

As a practical method to confirm entanglement of detectors system, the
test of Bell-CHSH~\cite{Clauser1969b} inequality for a pair of
detectors is usually accepted. In our previous
studies~\cite{nambu2011,Nambu2013}, we have confirmed that there is no
violation of Bell-CHSH inequality on the super horizon scale.
However, there is a possibility that effect of multipartite
entanglement can violate Bell-like inequalities.  Bell-Mermin-Klyshoko
(BMK) inequalities~\cite{Mermin1990a,Belinskii1993,Gisin1998} is such
a candidate which can capture multipartite entanglement. Several
authors discuss cosmological implication of this
inequalities~\cite{Kanno2017}. It may be interesting task to evaluate
degrees of violation of these inequalities for the present detectors
model.

\begin{acknowledgments}
  We would like to thank S.~ Ishizaka for informing us monogamy
  relation of negativity for qubit system. This work was supported in
  part by the JSPS KAKENHI Grant Number 16H01094.
\end{acknowledgments}


\end{document}